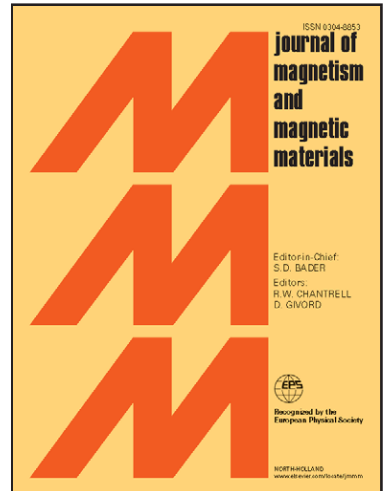

www.elsevier.com/locate/jmmm

# Magneto-Optic Faraday Effect in Maghemite Nanoparticles/Silica Matrix Nanocomposites prepared by the Sol-Gel Method

M. Domínguez[a,*], D. Ortega[b], J.S. Garitaonandía[c], R. Litrán[a], C. Barrera-Solano[a], E. Blanco[a] and M. Ramírez-del-Solar[a]

[a]*Dept. de Física de la Materia Condensada, Universidad de Cádiz, E11510 Puerto Real, Spain*
[b]*Dept. de Ciencia de los Materiales e Ingeniería Metalúrgica y Química Inorgánica, Universidad de Cádiz, E11510 Puerto Real, Spain*
[c]*Dept. de Física Aplicada II, Universidad del País Vasco, P.O. Box 644, E48080 Bilbao, Spain*

**Abstract**

Bulk monolithic samples of $\gamma$-$Fe_2O_3$/$SiO_2$ composites with different iron oxide/silica ratios have been prepared by the sol-gel technique. Iron oxide nanoparticles are obtained in-situ during heat treatment of samples and silica matrix consolidation. Preparation method was previously optimized to minimize the percentage of antiferromagnetic $\alpha$-$Fe_2O_3$ and parallelepipeds of roughly $2\times5\times12$ mm, with good mechanical stability, are obtained. RT magnetization curves show a non-histeretic behavior. Thus, magnetization measurements have been well fitted to an expression that combines the Langevin equation with an additional linear term, indicating that some of the nanoparticles are still superparamagnetic as confirmed by X-Ray Diffraction and Electron Microscopy measurements. ZFC/FC experiments show curves with slightly different shapes, depending on the size and shape distribution of nanoparticles for a given composition. Magneto-optical Faraday effect measurements show that the Faraday rotation is proportional to magnetization of the samples, as expected. As a demonstration of their sensing possibilities, the relative intensity of polarized light, measured at 5º from the extinction angle, was plotted versus applied magnetic field.

---

\* Corresponding author. Tel.: +34-956-016324; fax: +34-956-016288.
*E-mail address*: manolo.dominguez@uca.es

## 1. Introduction

Nanocomposites made up of particles embedded in a porous matrix have attracted considerable interest, since the structural confinement of the nanoparticles allows tailoring their physical properties [1, 2]. A porous matrix provides, at the same time, enough nucleation sites for nanoparticle formation and an effective way to avoid their aggregation. Silica gels are ideal for this application due to their chemical inertness, large accessible surface area, high porosity and transparency. If the final nanocomposite had the adequate properties, it would be a serious candidate for applications in magnetic field sensors. In general, magnetic materials are non transparent and Faraday rotation can only be measured in very thin films. However, a material consisting of magnetic nanoparticles in a transparent silica matrix allows measuring the Faraday rotation as the change in the polarization plane of the polarized light transmitted through the material, as a function of applied magnetic field.

## 2. Experimental

Monolithic nanocomposites consisting in $\gamma$-$Fe_2O_3$ (maghemite) nanoparticles embedded in a silica gel have been prepared using the traditional sol-gel method [3]. The silica matrices have excellent mechanical and optical properties and their pores allow the additional species to react in a restricted space. The starting materials were tetraethylorthosilicate (TEOS), water, ethanol and $Fe(NO_3)_3 \cdot 9H_2O$. The molar ratio TEOS:$C_2H_5OH$:$H_2O$ was 2:7:15. The TEOS was dissolved in ethanol using magnetic stirring for 10 min. An appropriate amount of $Fe(NO_3)_3 \cdot 9H_2O$ (for 3, 5 and 8 at% Fe/(Fe+Si)) was dispersed in a $HNO_3$ acidified pH 1 water and then, magnetically stirred to yield a final net composition with 4, 6.7 and 10.8 wt% iron oxide, respectively. Formamide, used as a drying control chemical to avoid cracking [4], was finally added in a molar ratio of 4 mol/mol TEOS. The samples were prepared at room temperature, sealed and placed at 50ºC under normal atmosphere. The time of gelation was approximately 90 min. Bulk pieces of around 0.1 cm$^3$ were thus obtained. After two days, the samples were washed in ethanol, sealed and uphold again at 50ºC for seven days. Next, the gels were partially uncovered and maintained at the same temperature for seven additional days to complete the drying. The dried gels were stored under vacuum at room temperature until thermal treatment was carried out. They were heated at 100ºC for 240 min, 350ºC for 400 min and 700ºC for 120 min in a nitrogen atmosphere (0.1 l/min) to remove the impurities of the pore network, ensure the crystallization process and stabilize them. The heating rate was 0.5ºC/min in all the cases.

The final composition was checked by Inductively Coupled Plasma-Atomic Emission Spectrometry (ICP-AES), using an Iris Intrepid HR spectrometer from ThermoElemental. X-Ray Diffraction (XRD) measurements were performed with a Bruker D8 Advance diffractometer in a Bragg-Brentano geometry configuration. Transmission Electron Microscopy has been performed in a JEOL 2010F microscope. RT magnetization measurements were made with a Faraday Balance (Oxford Instruments) while ZFC/FC curves were obtained with a Quantum Design SQUID magnetometer.

## 3. Results and discussion

Fig. 1 shows the XRD patterns for 5 and 8% samples. Peaks from tetragonal maghemite are marked. XRD pattern for the other composition is similar, but the peaks intensities are hardly visible in this case. In Fig. 2, a TEM micrograph of the 8% sample is shown where nanoparticles are clearly observed. Size of these particles ranges 5-15 nm. By HRTEM, it is possible to isolate a single particle (Fig. 2b). The growth mechanism of these particles has been simulated by the Bravais-Friedel-Donnay-Harker method, as shown in Fig. 2c.

Fig. 3 shows the room temperature magnetization for the three samples under study. The magnetic behavior is essentially anhisteretic, as corresponds to a set of superparamagnetic (SPM) nanosized maghemite particles. The magnetization curves are well fitted to a modified Langevin function that includes an additional linear term:

$$\sigma = \chi_g H + \sigma_s \left[ \coth\left(\frac{\mu_0 \mu H}{k_B T}\right) - \frac{k_B T}{\mu_0 \mu H} \right] \quad (1)$$

where $\sigma$ is the specific magnetization, $\chi_g$ is the mass susceptibility, $\sigma_s$ is the specific saturation magnetization, $\mu$ is the average magnetic moment of particles and $\mu_0$ and $k_B$ are the magnetic permeability of vacuum and the Boltzmann constant, respectively. The linear terms in eq. (1) may arise from the contribution of paramagnetic species, including impurities. It has been found in ferromagnetic resonance experiments [5] that a small amount of iron appears in the form of $Fe^{3+}$ ions, embedded in the silica network. Table 1 summarizes the data obtained in the fits of the three magnetization curves to eq.

(1). Taken into account the spontaneous magnetization of maghemite ($M_s$ = 370 kA/m), from the values of the average magnetic moment of particles, it is possible to estimate the size of such nanoparticles, if they are considered as spherical, from the relationship:

$$\mu = M_s \langle V \rangle = M_s \frac{\pi}{6} \langle D \rangle^3 \qquad (2)$$

Looking at the TEM micrographs, this assumption seems to be a good approximation. These estimated values for average particle diameter also appear in Table 1, and they are similar (around 12 nm) which is consistent with TEM micrographs. Obviously, a lower iron content means a lower number of maghemite nanoparticles (which is reflected in the term $\sigma_s$ of eq. (1) that equals to $N\mu$, $N$ being the number of particles per volume unit), but the average size of particles is not very different to what is seen in Fig. 2 for the 8% sample.

On the other hand, Fig. 4 shows the low field Zero Field Cooled (ZFC)/Field Cooled (FC) curves of the three samples. ZFC curves show the typical peak of superparamagnetic systems at the average blocking temperature (which corresponds to the temperature where the average size particle is blocked). The 8% sample has a wider peak, indicating that the distribution of particle sizes must be wider too.

According to the theory of superparamagnetism [6], the blocking temperature is related to the average particle volume:

$$T_B = \frac{K \langle V \rangle}{25 k_B} \qquad (3)$$

where $K$ is the effective uniaxial magnetic anisotropy constant and the value 25 is obtained when an experiment time of 100 s is considered. If the value of bulk maghemite is considered ($K = 2.5 \times 10^4$ J/m$^3$ [6]), $T_B \approx 66$ K (for a diameter of 12 nm), which is not too far from the values shown in ZFC curves. Thus, the magnetic anisotropy is not enlarged by additional contributions as dipolar interactions between particles or from shape anisotropy.

Finally, the magneto-optical characterization of the samples has been carried out. Specific Faraday rotation has been measured at 95.5 kAm$^{-1}$ for comparison between the samples (Table 1), with $\lambda$ = 680 nm, where the effect reaches a maximum. It ranges 69-95 rad/Tm on increasing the iron content. These values are of the same order of magnitude of the Verdet constant in commercial TGG (terbium gallium garnet, used in optical isolators) which has been reported to be around -134 rad/T·m at 633 nm. Fig. 5 shows the magneto-optical properties of the sample in a different way, by measuring the transmitted light intensity relative to the zero magnetic field value, when the polarizer/analyzer is rotated 5º from the extinction plane. This is a configuration used for sensing purposes and allows obtaining curves proportional to the Faraday rotation. They follow magnetization rather than applied magnetic field, as expected for a ferrimagnetic material as maghemite, at low magnetic field. At higher fields, when magnetization approaches saturation, the intensity curve deviates from scaling with magnetization. In fact, the insets in Fig. 5 show the plots of relative intensity vs. sample magnetization. They are linear only at low magnetization (i.e., at low applied field). This additional Faraday rotation, that is almost linear with field in this range, could be due to the effect of the doped amorphous silica matrix itself.

## 4. Conclusions

We have prepared $\gamma$-$Fe_2O_3$ nanoparticles/$SiO_2$ matrix composites, with several iron content, in the form of monolithic pieces, using the sol-gel method and carefully controlling preparation conditions. Electron Microscopy and magnetic measurements show that superparamagnetic nanoparticles are essentially spherical with average diameters around 12 nm. Magneto-optical characterization indicates that these materials have a specific Faraday rotation similar to the Verdet constant of materials used in commercial Faraday rotators and isolators. Further optimizations of the preparation process may allow increasing this value for low magnetic field, permitting to use them in magnetic sensing equipment.

**Acknowledgements**

This work has been financed by the Spanish MEC (MAT2002-02179) and by the Junta de Andalucia (PAI-FQM-335). The authors wish to thank R.P. Garcia for his assistance in magneto-optical setup and measurements.

**Table 1**. Summary of chemical analysis and magnetic data of the three samples under study (see text for details).

| Sample name | Nominal $Fe_2O_3$ content, wt% | Measured $Fe_2O_3$ content, wt% | $\chi_g$, m³/kg | $\sigma_s$, Am²/kg | $\mu_{np}$, Am² | $\langle D \rangle$, nm | Specific Faraday Rotation, rad/T·m |
|---|---|---|---|---|---|---|---|
| 3% | 4.0 | 3.6 | $4.0 \times 10^{-9}$ | 1.85 | $3.2 \times 10^{-19}$ | 11.6 | 69.5 |
| 5% | 6.7 | 5.9 | $9.2 \times 10^{-9}$ | 3.12 | $3.6 \times 10^{-19}$ | 12.1 | 83.3 |
| 8% | 10.8 | 9.6 | $1.4 \times 10^{-8}$ | 3.81 | $3.3 \times 10^{-19}$ | 11.7 | 94.4 |

**Figure Captions**

Fig. 1. X-Ray Diffraction Pattern for 8% and 5% samples. Dots indicate the positions of reflections that correspond to tetragonal maghemite.

Fig. 2. (a) TEM micrograph of an 8% sample. Arrows are indicating the position of several nanoparticles; (b) a HRTEM image of an isolated nanoparticle and (c) a model of the nanoparticle based on the Bravais-Friedel-Donnay-Harker method.

Fig. 3. Room Temperature Magnetization curves for the samples under study. The lines correspond to the fits to eq. (1). The values of the fit parameters deduced here appear in Table 1.

Fig. 4. ZFC/FC curves for the three samples under study. All measurements were made at 3.98 kAm$^{-1}$ (50 Oe).

Fig. 5. Relative polarized light intensity versus applied magnetic field (the line is a guide to the eye). The insets show the same data plotted as a function of the sample magnetization, deduced from the fit to eq. (1).

Fig. 1

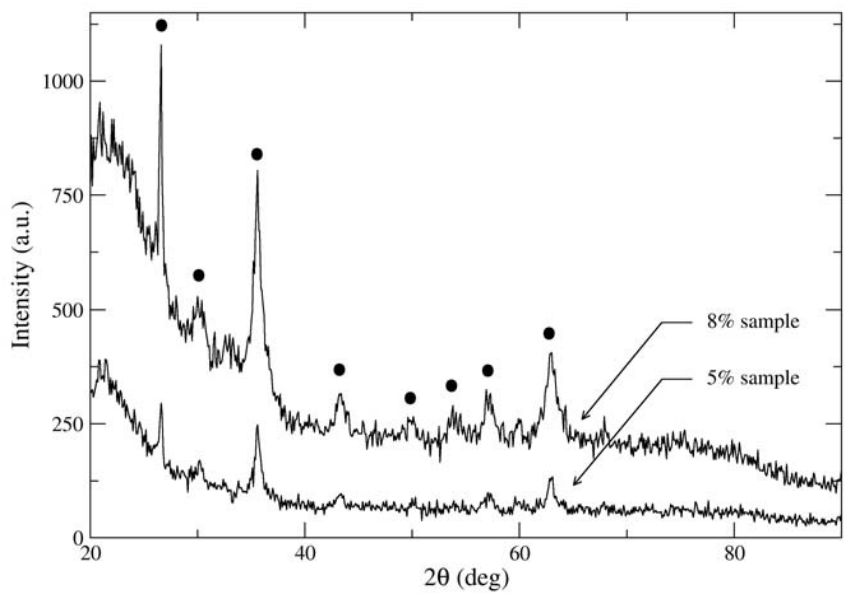

Fig. 2

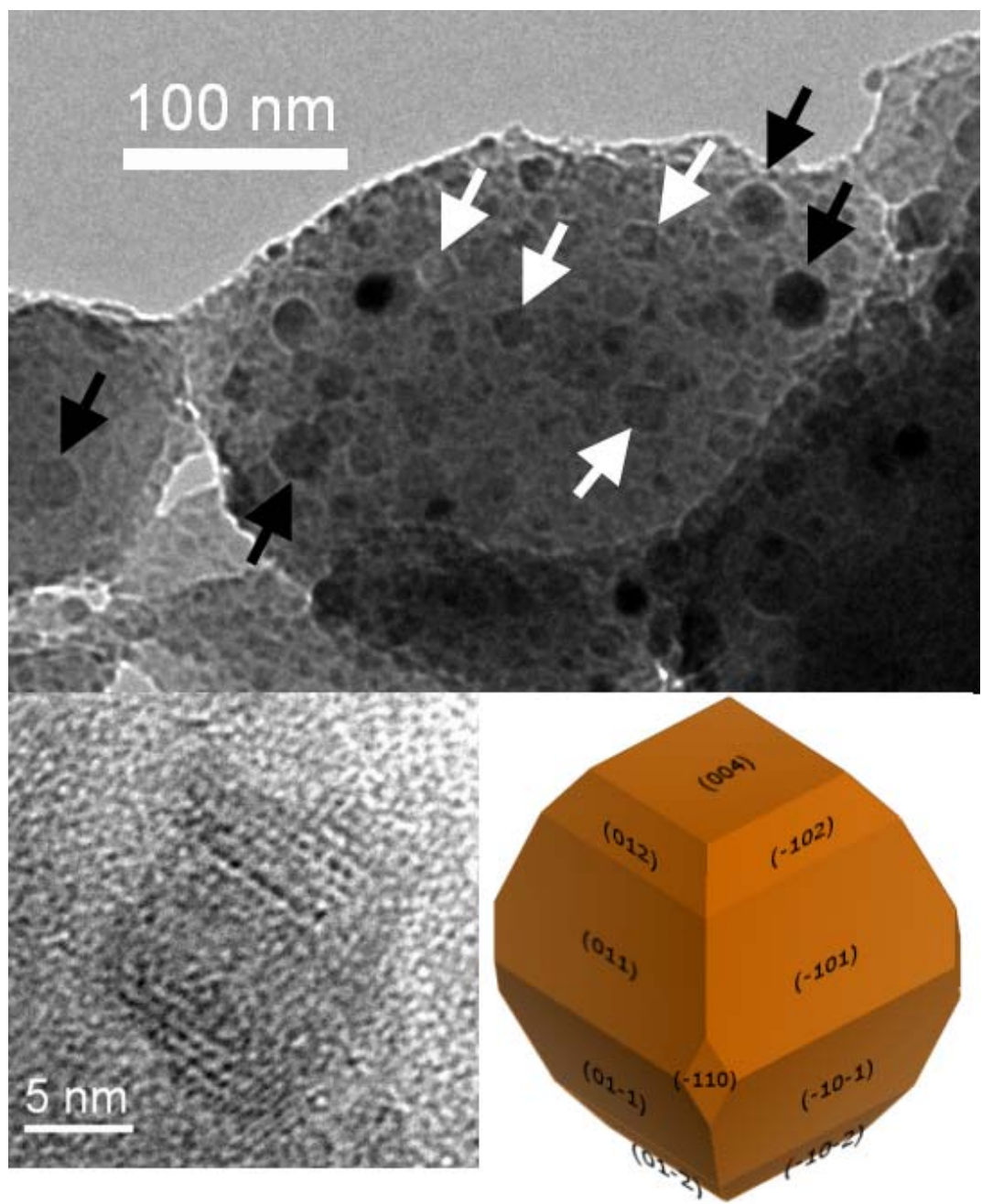

Fig. 3

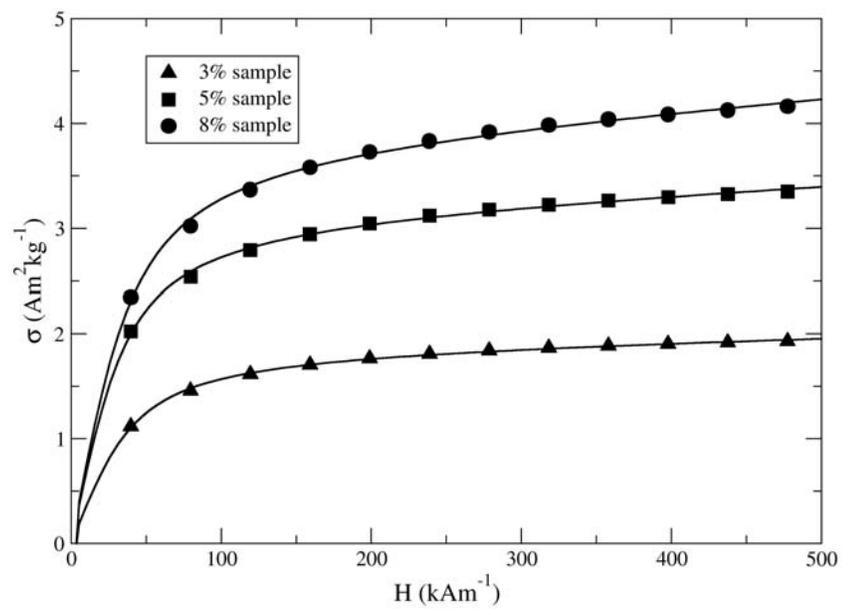

Fig. 4a

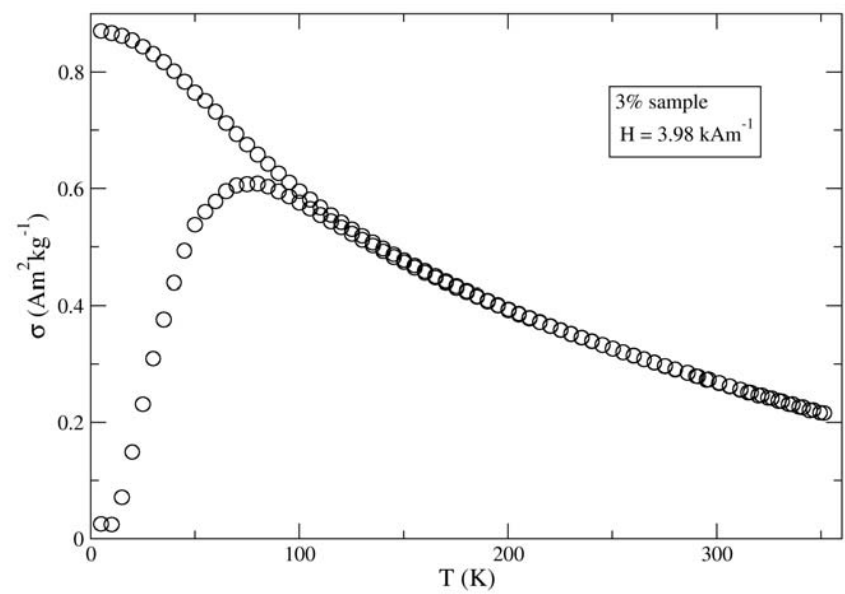

Fig. 4b

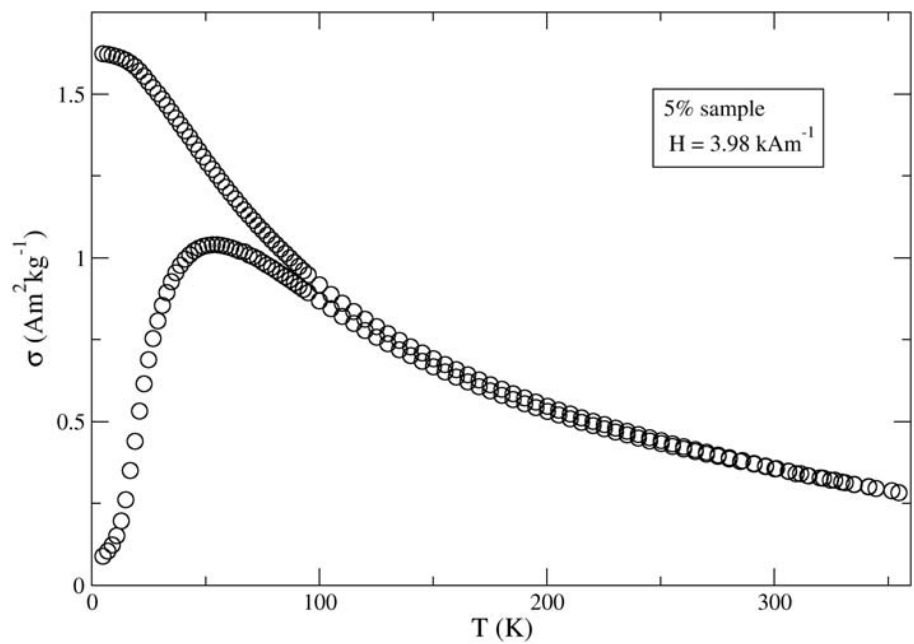

Fig. 4c

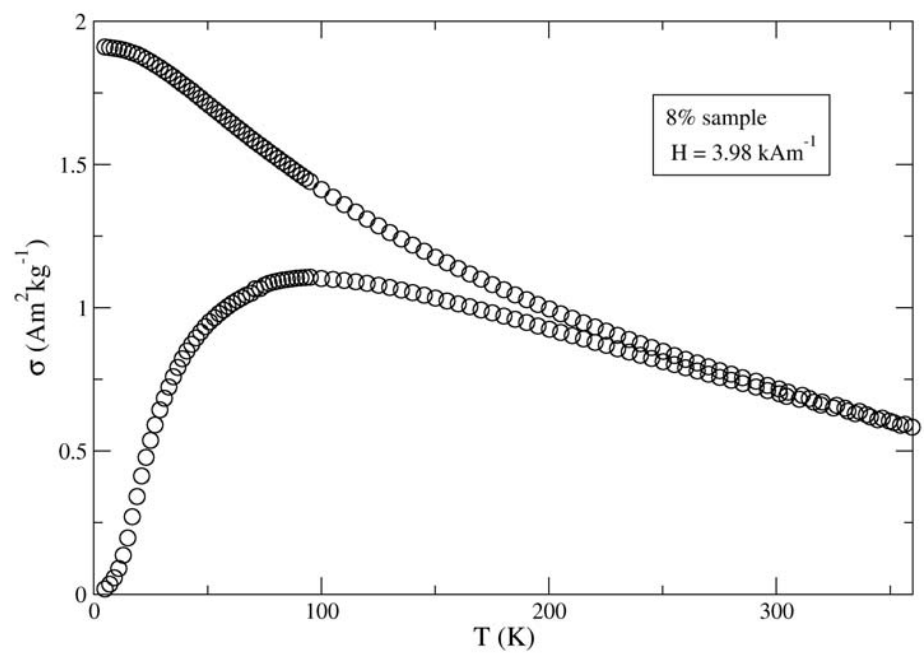

Fig. 5a

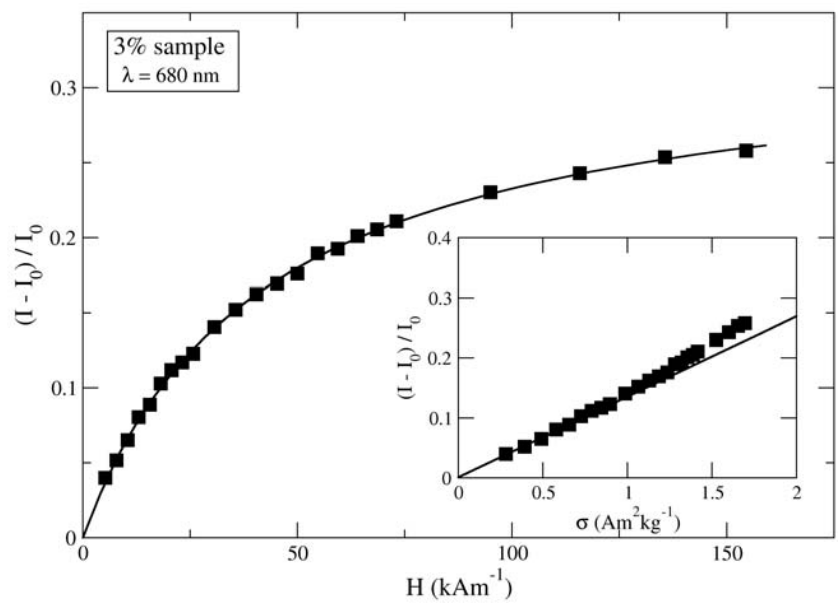

Fig. 5b

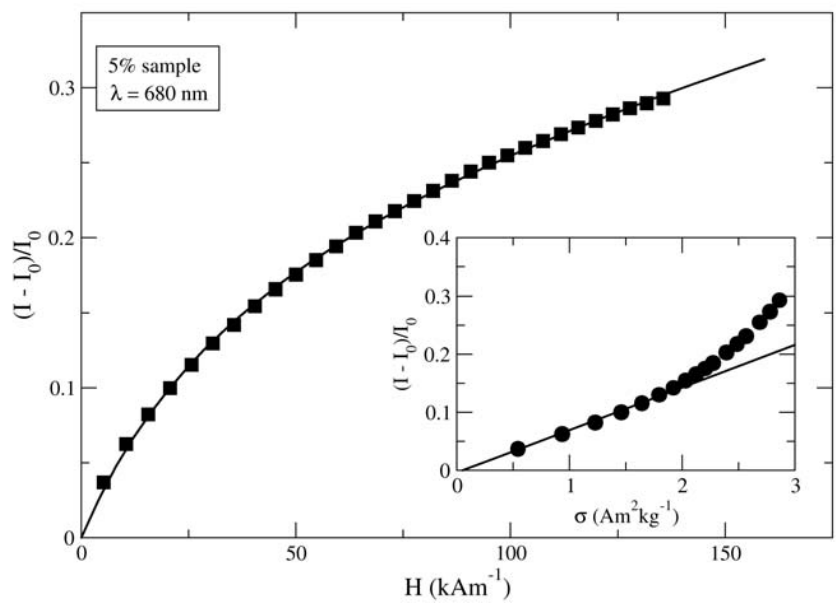

Fig. 5c

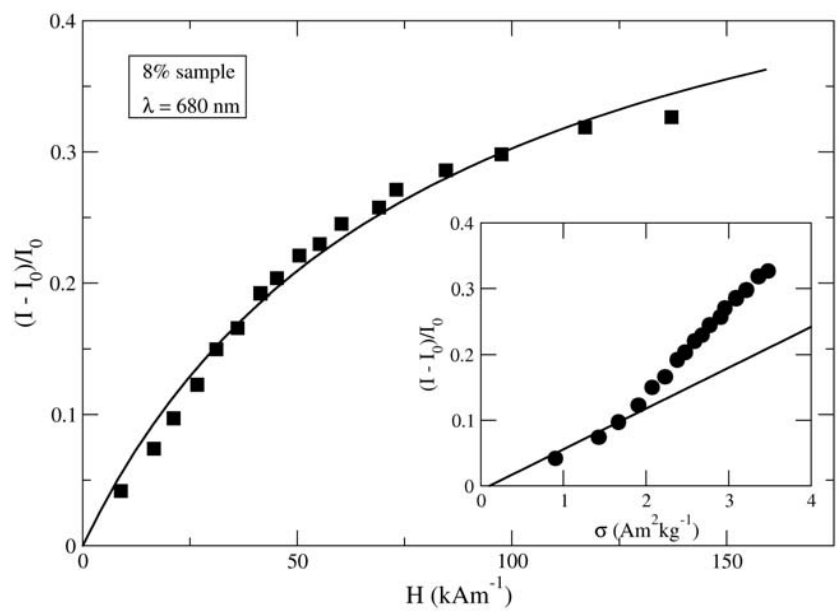